\documentclass[a4paper]{jpconf}
\usepackage{graphicx}
\usepackage{subfigure}
\usepackage{bm}
\usepackage{amsmath}

\begin{document}
\title{Short-range correlations in modified planar rotator model}

\author{M \v{Z}ukovi\v{c}\textsuperscript{1} and D T Hristopulos\textsuperscript{2}}

\address{\textsuperscript{1} Institute of Physics, Faculty of Science, P.J. \v{S}af\'arik University, Park Angelinum 9, 041 54 Ko\v{s}ice, Slovakia}

\address{\textsuperscript{2} Geostatistics Laboratory, School of Mineral Resources Engineering, Technical University of Crete,
Chania 73100, Greece}

\ead{milan.zukovic@upjs.sk}

\begin{abstract}
We introduce a model inspired from statistical physics that is shown to display flexible short-range spatial correlations which are potentially useful in geostatistical modeling. In particular, we consider a suitably modified planar rotator or XY model, traditionally used for modeling continuous spin systems in magnetism, and we demonstrate that it can capture spatial correlations typically present in geostatistical data. The empirical study of the spin configurations produced by Monte Carlo simulations at various temperatures and stages in the nonequilibrium regime shows that their spatial variability can be modeled by the flexible class of Mat\'{e}rn covariance functions. The correlation range and the smoothness of these functions vary significantly in the parameter space that consists of the temperature and the simulation time. We briefly discuss the potential of the model for efficient and automatic prediction of spatial data with short-range correlations, such as commonly encountered in geophysical and environmental applications.
\end{abstract}

\section{Introduction}
Modern remote sensing techniques allow recording enormous amounts of spatial
data, including geographical, natural resources, land use, and environmental remote sensing images, which need efficient (preferably real-time) processing~\cite{Rossi94,Foster08}. Such processing involves the reconstruction of missing data that often occur due to different reasons, such as equipment malfunctions and gaps in the coverage of the targeted area that appear as a result of restricted satellite paths or bad weather conditions~\cite{Jun04,Lehman04,Albert12,Bechle13,Emili11}. However, such massive data sets can not be efficiently handled by standard geostatistical methods, such as kriging~\cite{Wack03}. The main drawbacks of such methods are high computational complexity, difficulties to automatize the algorithms to work without subjective user inputs (selection of variogram and search radius for interpolation) and often the necessity of some data pre-processing (e.g., lognormal transformation) if the data does not comply with the Gaussianity requirement~\cite{Dig07}. Thus, there is a need to develop new spatial prediction techniques that overcome these shortcomings~\cite{Cressie08,Hartman08}.
Recently, we have proposed efficient spatial classification methods based on models inspired from statistical physics, in particular discrete spin Ising, Potts, and clock models, employing a heuristic ``energy matching'' principle~\cite{zuk09a,zuk09b}.

In the present study we extend the idea of using spin models for spatial prediction purposes and introduce a new method that is based on a continuous spin planar rotator model. Even though the above mentioned nonparametric discrete-spin-based classification models have been shown to be rather competitive, their performance was based on the assumption of the existence of relevant correlations at some unknown parameters and the prediction results were discrete values even for continuous data. As we empirically demonstrate, the modified planar rotator model {\it inherently} displays flexible short-range spatial correlations that vary significantly over the model's parameter space and could be used for spatial interpolation of continuous geostatistical data.

\section{Model and simulations}

\subsection{Standard planar rotator model}
The Hamiltonian of the standard two-dimensional planar rotator model with nearest-neighbor interactions on a square lattice is defined as
\begin{equation}
\label{Hamiltonian}
H=-J\sum_{\langle i,j \rangle}{\bm s}_{i}{\bm s}_{j}=-J\sum_{\langle i,j \rangle}\cos(\phi_{i}-\phi_j),
\end{equation}
where ${\bm s}_i=(\cos\phi_i,\sin\phi_i)$ is a continuous spin on $i$-th lattice site, represented by a two-dimensional unit vector, $\phi_i \in [0,2\pi]$ is an angle associated with the spin ${\bm s}_i$, $J$ is an exchange interaction parameter and $\langle i,j \rangle$ denotes the sum over nearest neighbors. The Mermin-Wagner theorem~\cite{merm66} prevents any long-range ordering at finite temperatures in such a model, which has, however, been intensively studied in connection with quasi long-range ordering (the so called Kosterlitz-Thouless phase) that appears at low temperatures~\cite{kost73,kost74}. The phase transition is produced by the unbinding of vortex-antivortex pairs at the Kosterlitz-Thouless critical temperature $T_{\rm KT}$, below which all spins are almost aligned even though true long-range order is destroyed by spin fluctuations. The quasi long-range ordering for $T<T_{\rm KT}$ is characterized by the power-law decaying correlation function
\begin{equation}
\label{cor_fun}
C(h;\eta)= \langle {\bm s}(h){\bm s}(0) \rangle = \Big(\frac{h}{L}\Big)^{-\eta(T)},
\end{equation}
where $h$ is the spin separation distance, $L$ is the linear lattice size,  and $\eta(T)=T/2\pi J$ is the temperature-dependent exponent.

\subsection{Modified planar rotator model}
\begin{figure}[t]
\centering
\subfigure{\includegraphics[scale=0.5,clip]{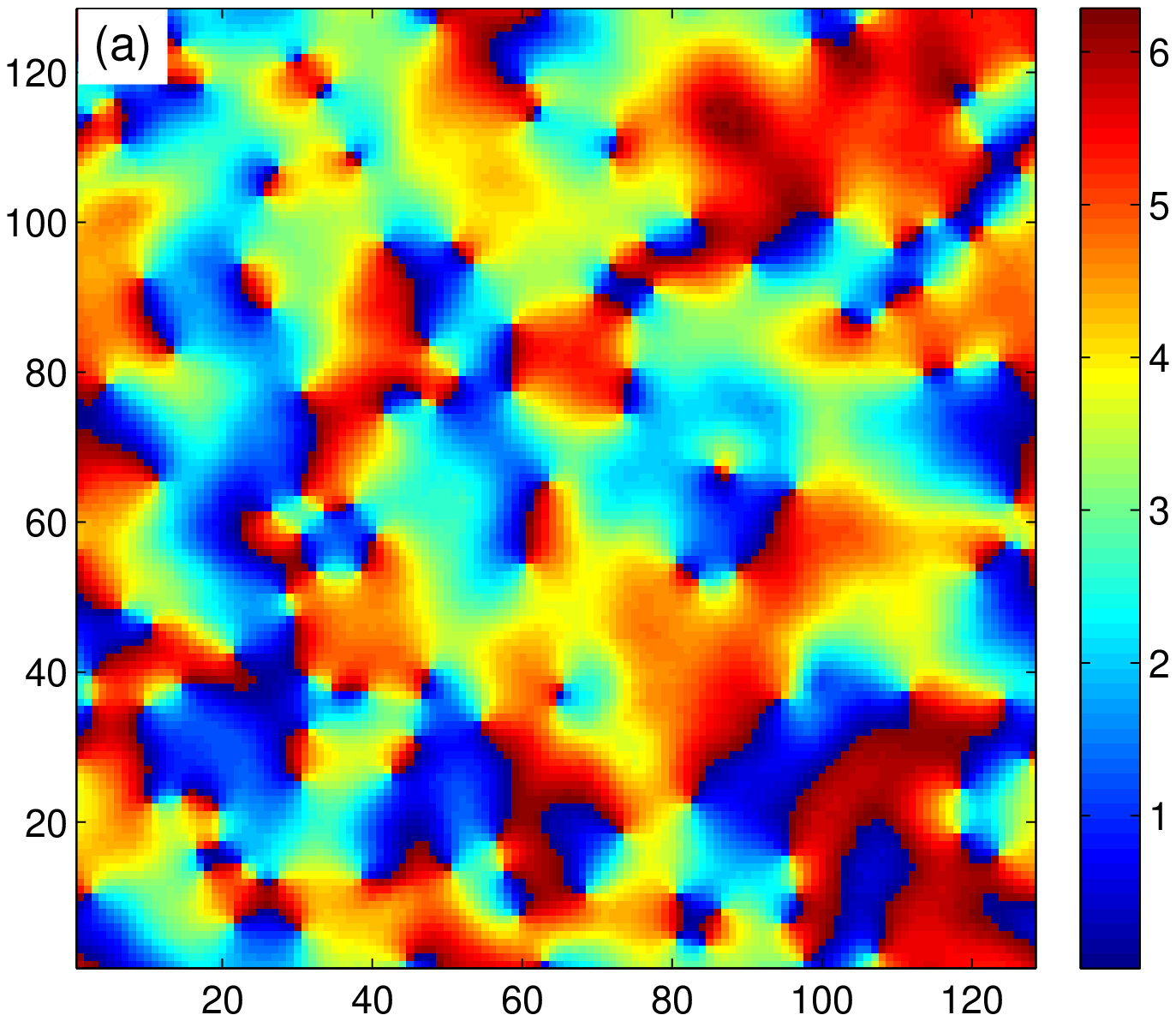}\label{fig:p1}}\hspace*{5mm}
\subfigure{\includegraphics[scale=0.5,clip]{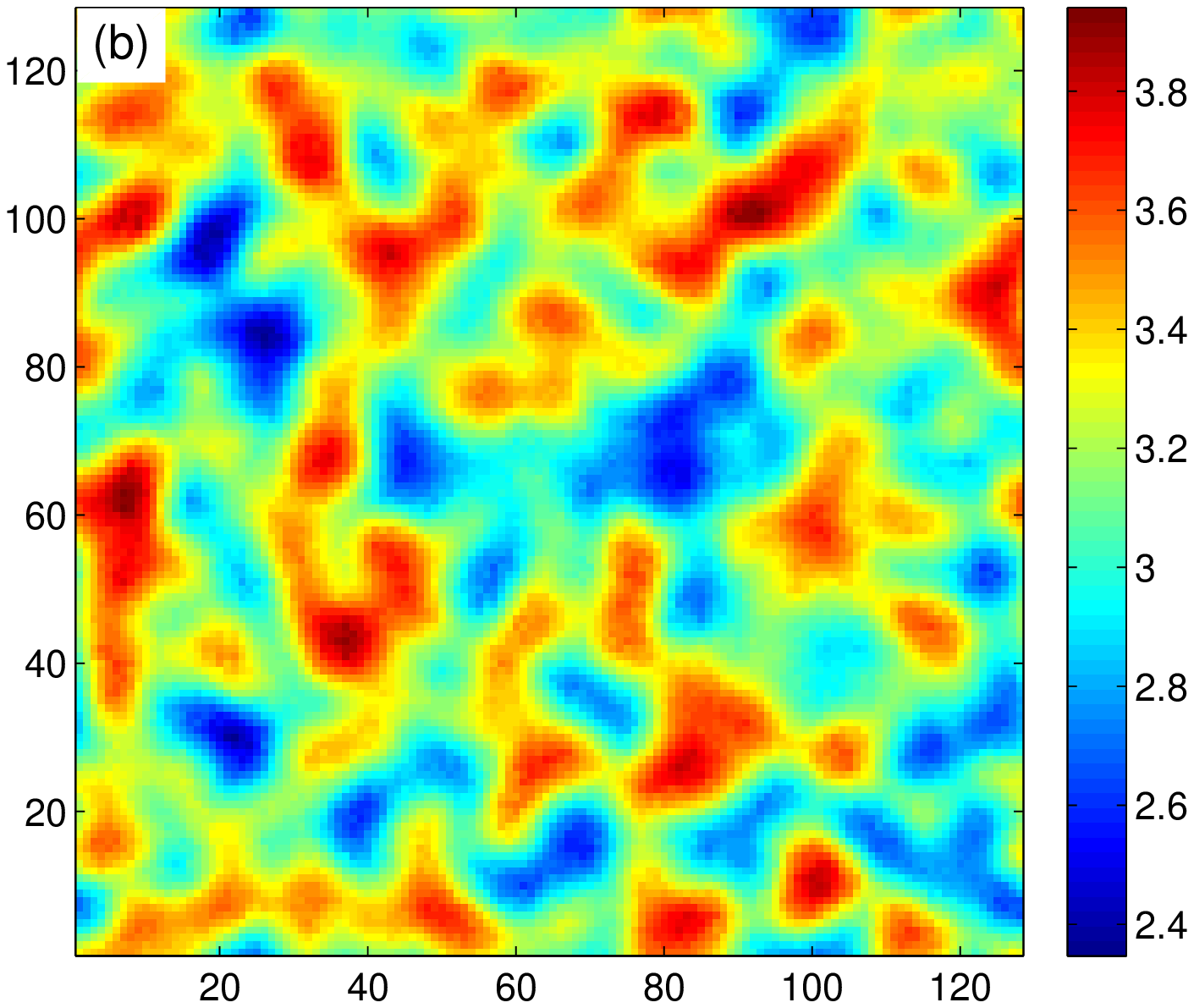}\label{fig:p1_2}}
\caption{Snapshots of spin configurations in the low-temperature regime ($T=10^{-4}$) for (a) standard and (b) modified planar rotator model, with periodic boundary conditions.}\label{fig:snaps}
\end{figure}

Even though the algebraic correlations with the tunable exponent $\eta(T)$ could be relevant for modeling of some spatial processes, there are some deficiencies that prevent the straightforward use of the standard planar rotator model for such purposes. First of all, one needs to define a proper mapping between the spin values and the geostatistical values. Geostatistical data are often positively correlated reflecting a degree of spatial continuity, which means that neighbors with similar values are more likely (i.e., have lower energy) than those with different values. Apparently, this condition is not fulfilled in the standard planar rotator model, described by the Hamiltonian~(\ref{Hamiltonian}); the latter, for example, assigns equal energies to a pair of neighbors whose angles differ by $\theta$ as to a pair whose angles differ by $2\pi+\theta$. This degeneracy could be fixed by defining the energy functional so that it monotonically increases with the turn angle between the neighboring spins on the entire interval $[0,2\pi]$. The latter can be achieved by modifying the Hamiltonian~(\ref{Hamiltonian}) to take the following form of the modified planar rotator (MPR) model
\begin{equation}
\label{Hamiltonian_mod}
H'=-J\sum_{\langle i,j \rangle}\cos[q(\phi_{i}-\phi_j)],
\end{equation}
where $0 < q \leq 1/2$ is a modification factor. In the following we will use a fixed value of $q = 1/2$. The difference between the nature of spatial correlations in the standard and modified models is apparent from snapshots of spin configurations (turn angles) in the low-temperature region, shown in Fig.~\ref{fig:snaps}. While the presence of the vortex-antivortex pairs shows up in Fig.~\ref{fig:p1} in the form of sharp boundaries between the domains of similarly oriented spins, the variation of the spin values in the MPR model, shown in Fig.~\ref{fig:p1_2} is rather smooth, as it is typical for geostatistical data.
\subsection{Monte Carlo simulations}
We use Monte Carlo (MC) simulations with the Metropolis update rule and vectorized checkerboard algorithm. The results are presented for a square lattice with the size $L \times L$. We chose $L=128$ as a compromise value between the computational resources needed for MC simulations and calculations of the empirical variograms (see below) on one side and the effort to secure ergodic conditions on the other side. In typical simulations of magnetic systems one would like to suppress boundary effects by employing periodic boundary conditions (see the snapshots in Fig.~\ref{fig:snaps}). However, these are not appropriate for simulation of geostatistical data, since normally there is no reason to assume the data on the opposite boundaries are correlated. Free boundary conditions are not appropriate either since the spatial process is not necessarily confined within the domain boundaries. Instead, we consider it reasonable to apply boundary conditions that assume the smooth continuation of the spatial process beyond the borders. This assumption is implemented by inserting additional nodes that decorate the lattice and requiring that these additional points have the same values as their nearest neighbors inside the lattice. Therefore, if ${\bm s}_{i,j}$ is a spin in the $i$-th row and $j$-th column of the lattice, $i,j=1,\dots L$, then ${\bm s}_{i,L+1}={\bm s}_{i,L}$, ${\bm s}_{L+1,j}={\bm s}_{L,j}$, ${\bm s}_{i,0}={\bm s}_{i,1}$ and ${\bm s}_{0,j}={\bm s}_{1,j}$, where the rows and columns with  indices $0$ and $L+1$ respectively are formed by auxiliary nodes around the lattice added to deal with the boundary effects. 

\subsection{Modeling spatial variability}
The MC simulations produce spatially correlated spin realizations ${\bm s}_{i}$, which can be represented by their turn angles $\phi_i$. It turns out that there is great flexibility in the spatial correlations as we move in the temperature and simulation time parameter space. Therefore, in order to model the spatial variability (the local variogram) of the simulated spin values we use a very flexible Mat\'{e}rn covariance model. In addition to the trivial parameters of the standard models (Gaussian, exponential, spherical, etc.), controlling the variance, $\sigma^{2}$, and the characteristic covariance length, $\xi$, the Mat\'{e}rn model involves one more parameter, $\nu$, controlling the smoothness of the spatial process. Therefore, it can be considered appropriate for those geostatistical applications in which controlling the smoothness is important. Furthermore, the Mat\'{e}rn model includes the Gaussian and the exponential models as special cases for $\nu \rightarrow \infty $ and $\nu=0.5$, respectively, as well as several other bounded models~\cite{Minasny05}. Due to its flexibility, it has been used in various areas including pedology~\cite{Minasny05,Minasny07,Marchant07}, hydrology~\cite{Rodri74,Pardo09}, topography~\cite{Handcock93}, health modeling~\cite{kamm03}, meteorology~\cite{Handcock94} and environmental modeling~\cite{fuentes02}. The Mat\'{e}rn correlation function has the general form
\begin{equation}
\label{mate_cf} C(h;{\bm \theta'})=\frac{{2}^{1-\nu}}{\Gamma(\nu)}\Big(\frac{h}{\xi}\Big)^{\nu}K_{\nu}\Big(\frac{h}{\xi}\Big),
\end{equation}
where ${\bm \theta'}=(\xi,\nu)$ and $K_{\nu}$ is the modified Bessel function of order $\nu$. Then the corresponding variogram function $\gamma(h;{\bm \theta})$, which is typically used in geostatistics, is related to the correlation function as
\begin{equation}
\label{mate_var} \gamma(h;{\bm \theta})=\sigma_n^{2}+\sigma^{2}\left[1-C(h;{\bm \theta'}) \right],
\end{equation}
where $\sigma^{2}$ is the random field variance, $\sigma_n^{2}$ is the nugget variance that corresponds to uncorrelated fluctuations and ${\bm \theta}=(\sigma_n,\sigma,\xi,\nu)$ is a vector of the complete model parameters.

Having generated realizations of spin values from MC simulations, we can assess their spatial variability by calculating the experimental (i.e., sample-based) variogram as
\begin{equation}
\label{emp_var} \hat{\gamma}(h)=\frac{1}{2N(h)} \sum_{i=1}^{N(h)}\left [\phi(x_i) - \phi(x_i+h) \right]^2,
\end{equation}
where $N(h)$ is the number of pairs of spins separated by the distance $h$.

There are several methods of fitting the model to the experimental variogram. In the present study we used the weighted least squares estimator (hereafter WLS) that was reported to give the best overall results~\cite{zimm91}. The WLS estimator is based on minimizing the weighted sum of squared residuals (objective function), which is defined as follows:
\begin{equation}
\label{wls} O({\bm \theta})= \sum_{k=1}^{k_{\max}}
\frac{N(h_k)}{[\gamma(h_k;{\bm \theta})]^2}[\hat{\gamma}(h_k)-\gamma(h_k;{\bm \theta})]^2,
\end{equation}
where $\hat{\gamma}(h_k)$ is the experimental and $\gamma(h_k;{\bm \theta})$ the model variogram. The summation runs over the $k=1,\ldots,k_{\max}$ lags containing $N(h_{k})$ pairs of points, where $h(k_{\max})$ is less than one half of the maximum lag in the sample.

\section{Results and discussion}
\begin{figure}[t!]
\subfigure{\includegraphics[scale=0.54,clip]{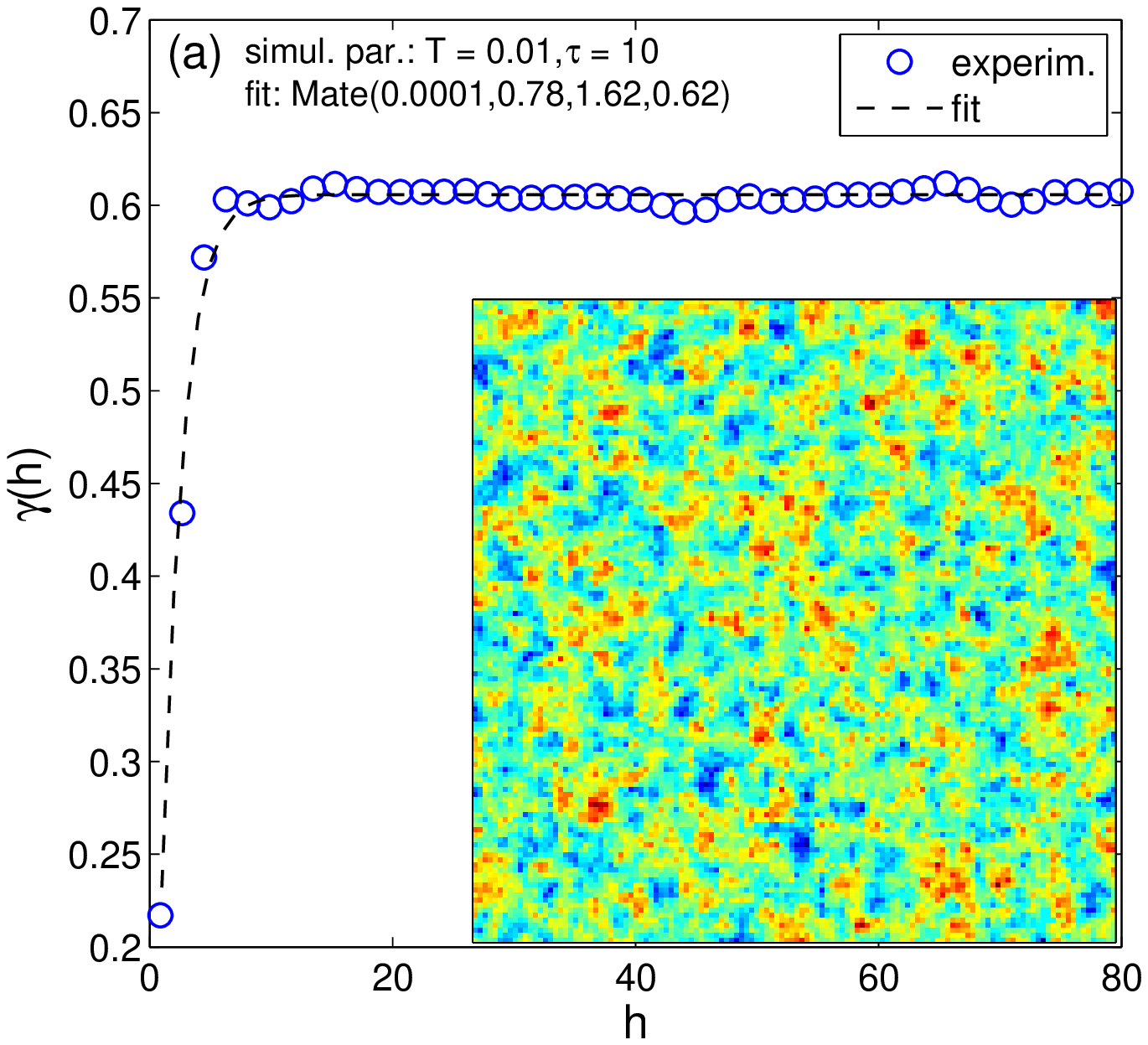}\label{fig:t001_mcs10}}
\subfigure{\includegraphics[scale=0.54,clip]{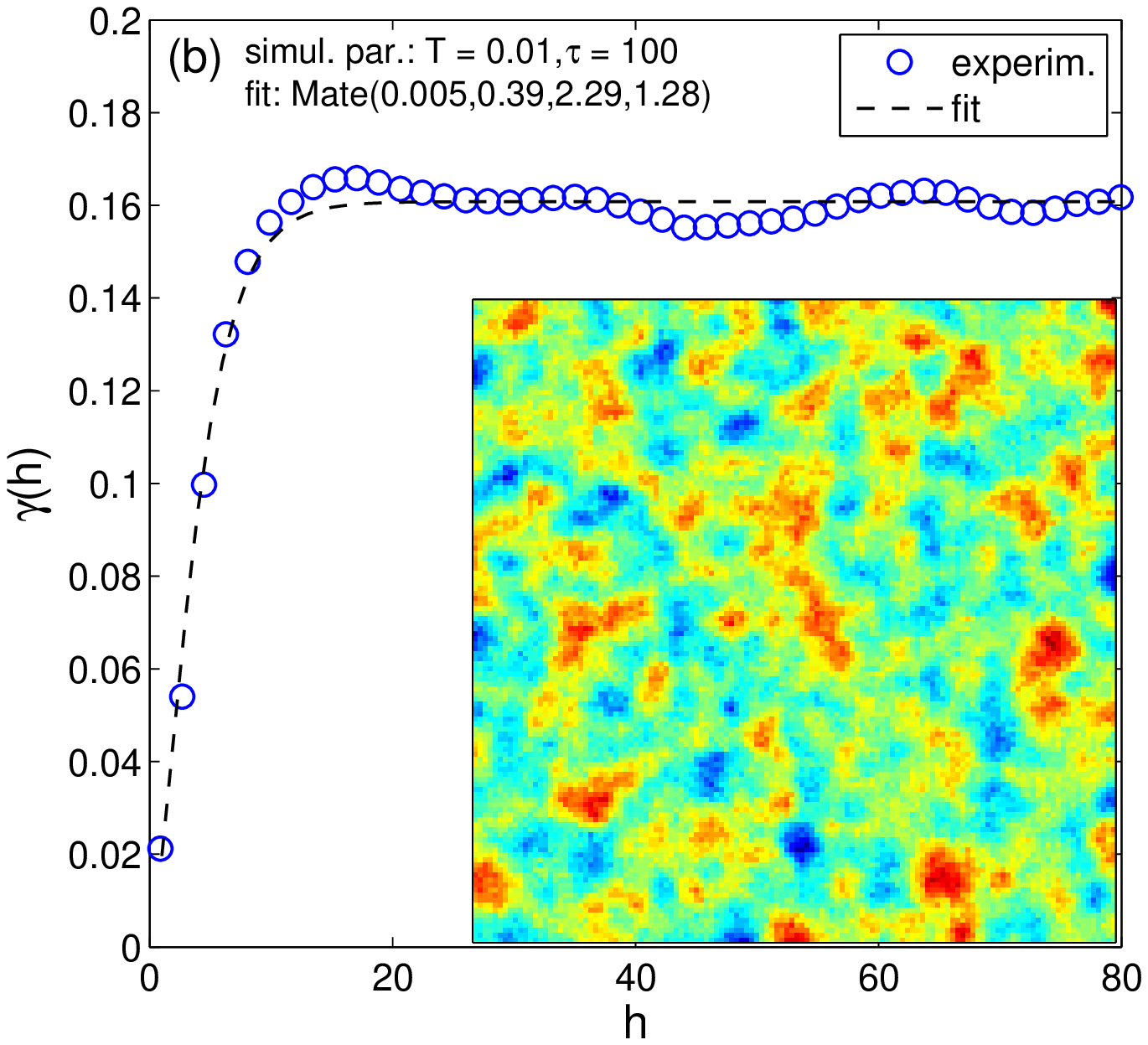}\label{fig:t001_mcs100}}
\subfigure{\includegraphics[scale=0.54,clip]{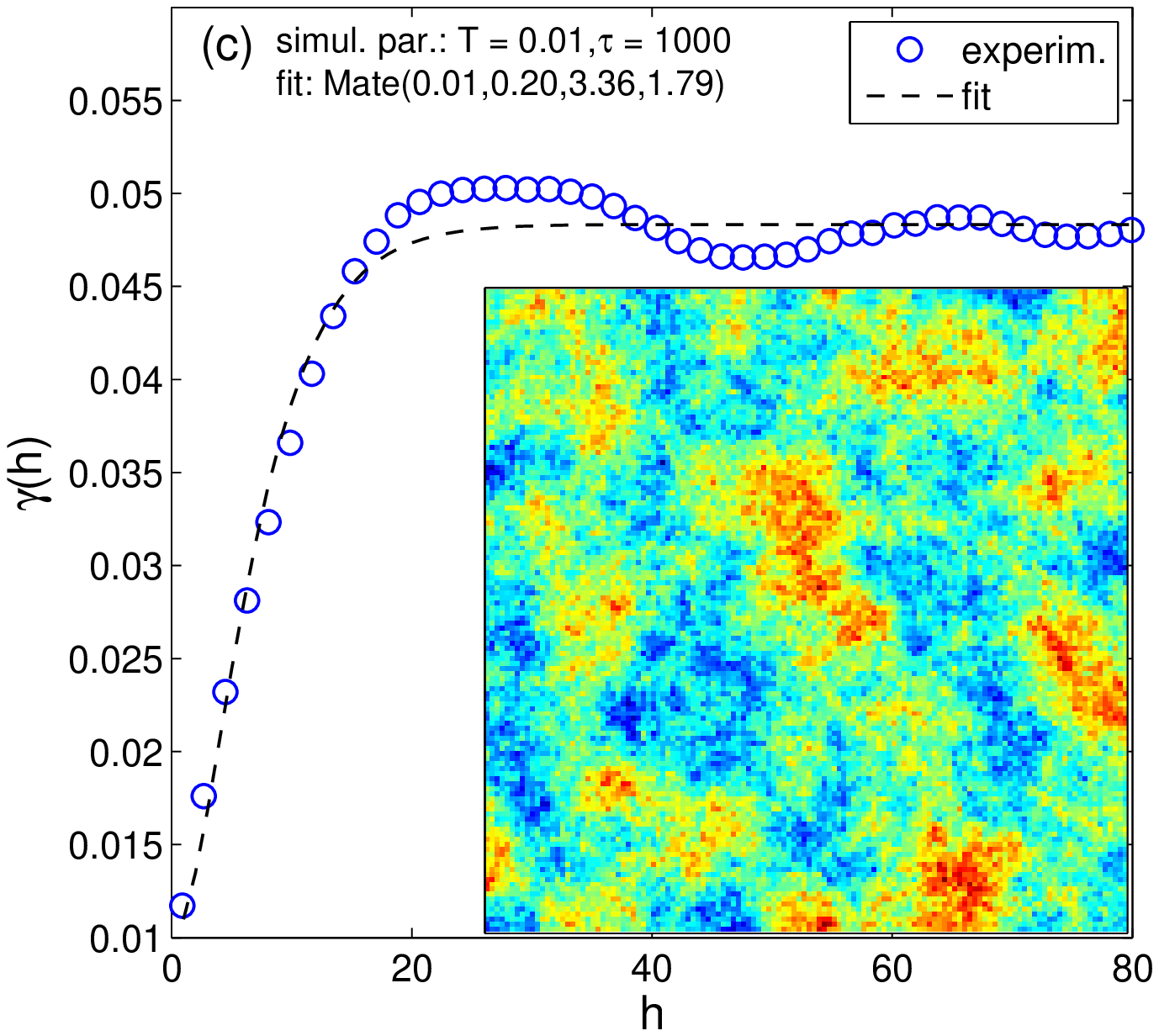}\label{fig:t001_mcs1000}}
\subfigure{\includegraphics[scale=0.54,clip]{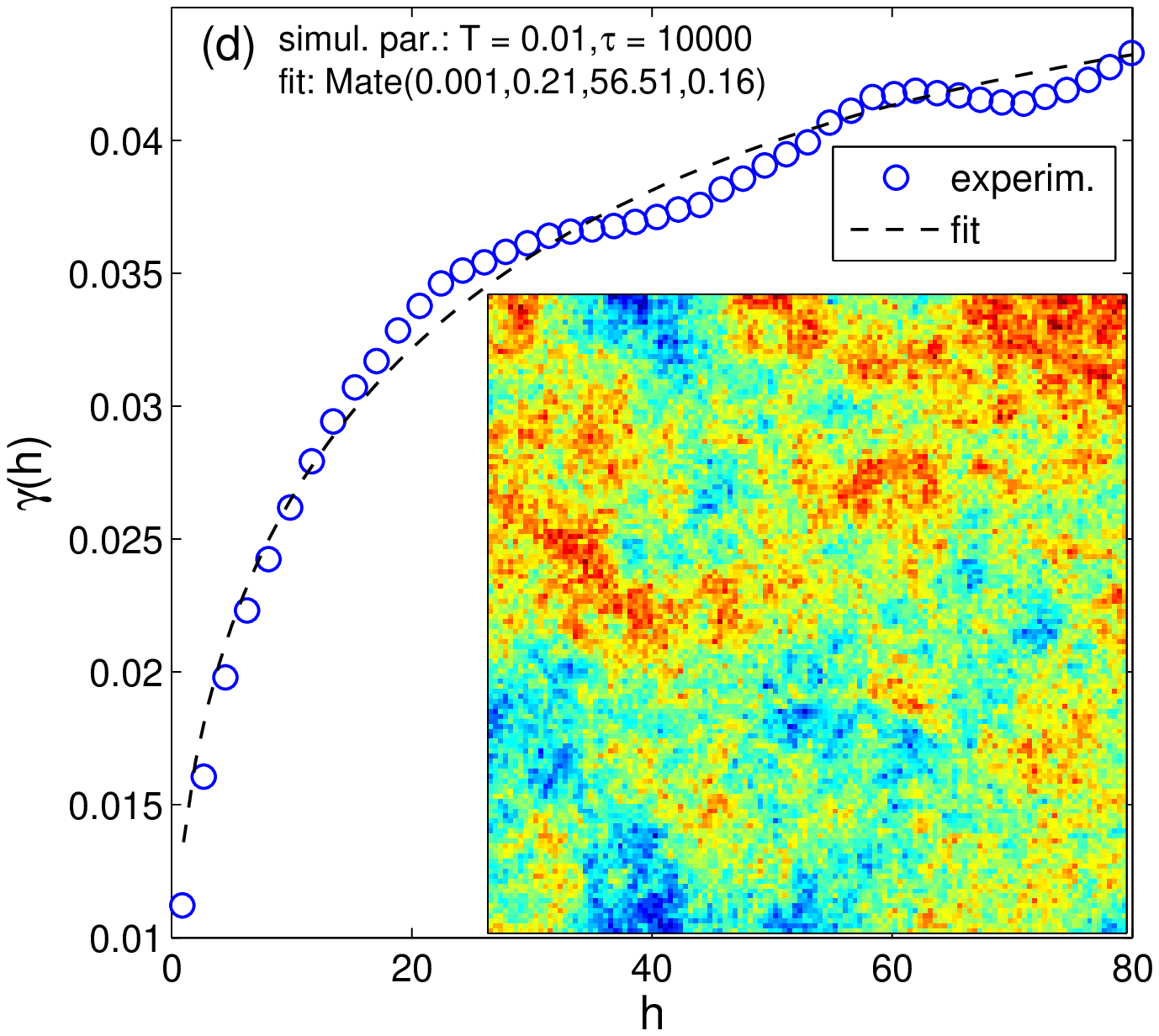}\label{fig:t001_mcs10000}}
\caption{\label{fig:T001}Experimental and fitted variograms of spin realizations ---shown in the inset--- obtained at temperature $T=10^{-2}$ and at various simulation times $\tau$ equal to (a) 10, (b) 100, (c) $10^3$ and (d) $10^4$ MC sweeps. Mate($\hat{\bm \theta}$) denotes the Mat\'{e}rn model for the inferred parameter values $\hat{\bm \theta}=(\hat{\sigma}_n,\hat{\sigma},\hat{\xi},\hat{\nu})$.}
\end{figure}

In order to empirically study spatial correlations in the realizations of spin values generated by the modified planar rotator model we run MC simulations at various temperatures $T$ (in units $J/k_B$, where $k_B$ is Boltzmann constant). Each simulation starts from an initial configuration of spatially uncorrelated random spin angles uniformly distributed in the interval $[0,2\pi]$ and the snapshots are collected after $\tau$ MC sweeps. Then, using Eq.~(\ref{emp_var}) the experimental variogram is calculated and subsequently fitted to the appropriate model by the WLS method defined by Eq.~(\ref{wls}). The results obtained at relatively higher temperatures are demonstrated in Fig.~\ref{fig:T001}, for $T=10^{-2}$ and $\tau=10$, 100, 1000 and 10000. Soon after the start of the simulation one can observe the development of spatial correlations both in the snapshots in the form of nucleation of small spin domains with similar values as well as in the small-scale behavior of the experimental variograms. The fitted parameters $\hat{\xi}=1.62$ and $\hat{\nu}=0.62$ for $\tau=10$ in Fig.~\ref{fig:t001_mcs10} indicate that the realization is quite rough with a rather small characteristic length. As the relaxation proceeds the variance decreases and both the characteristic length and the smoothness parameter increase, as shown in Figs.~\ref{fig:t001_mcs100} and~\ref{fig:t001_mcs1000}, for $\tau=100$ and 1000, respectively. Nevertheless, as equilibrium is approached the realizations become rougher again, whereas the characteristic length further increases and saturates only in equilibrium to a temperature-dependent value. The estimated characteristic length $\hat{\xi}=56.51$ for the equilibrium configuration at $\tau=10000$ in Fig.~\ref{fig:t001_mcs10000} implies that the samples on the $128 \times 128$ lattice suffer from lack of ergodicity, which is also reflected in the trend displayed by the corresponding experimental variogram.

\begin{figure}[t!]
\subfigure{\includegraphics[scale=0.54,clip]{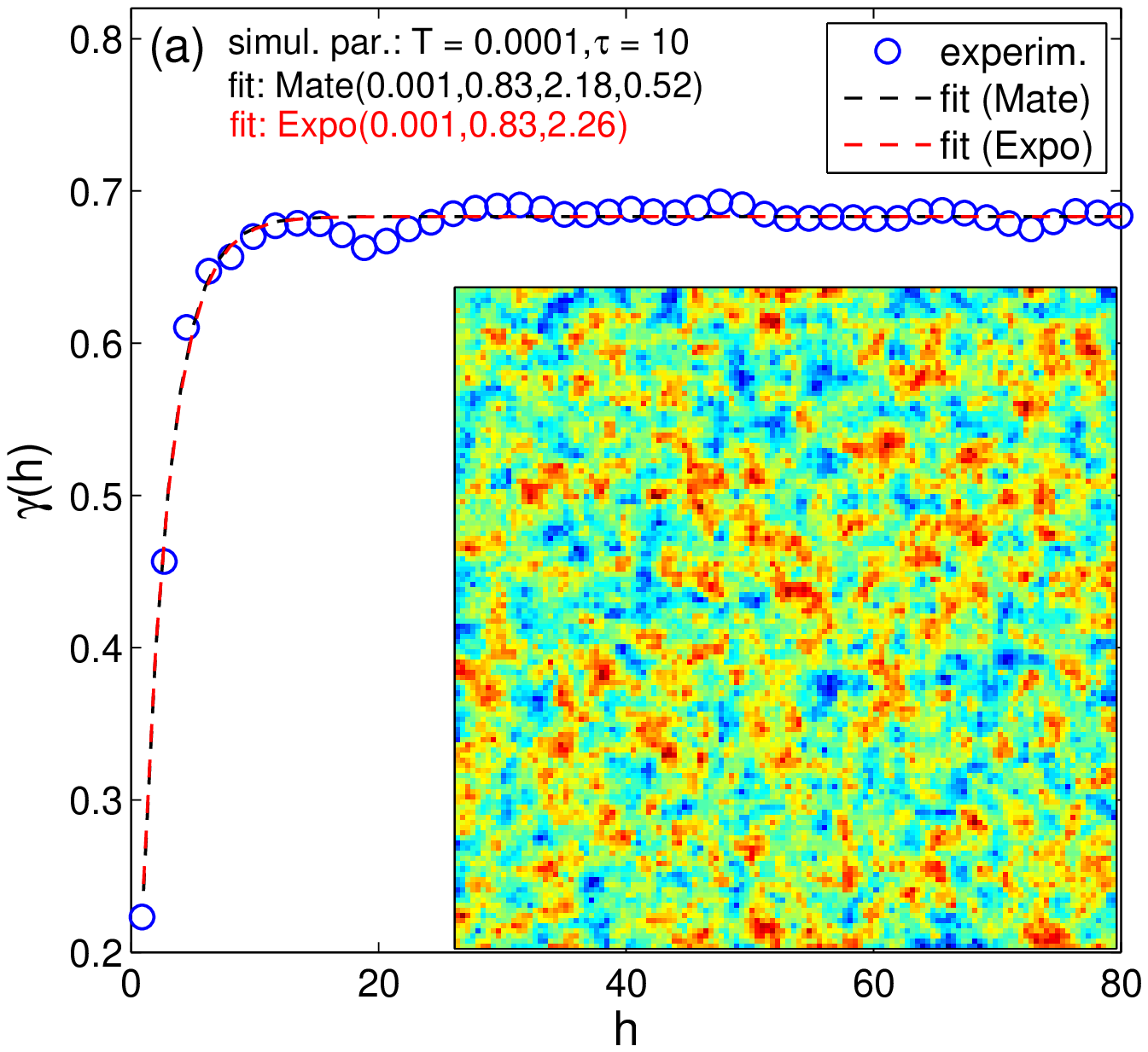}\label{fig:t00001_mcs10}}
\subfigure{\includegraphics[scale=0.54,clip]{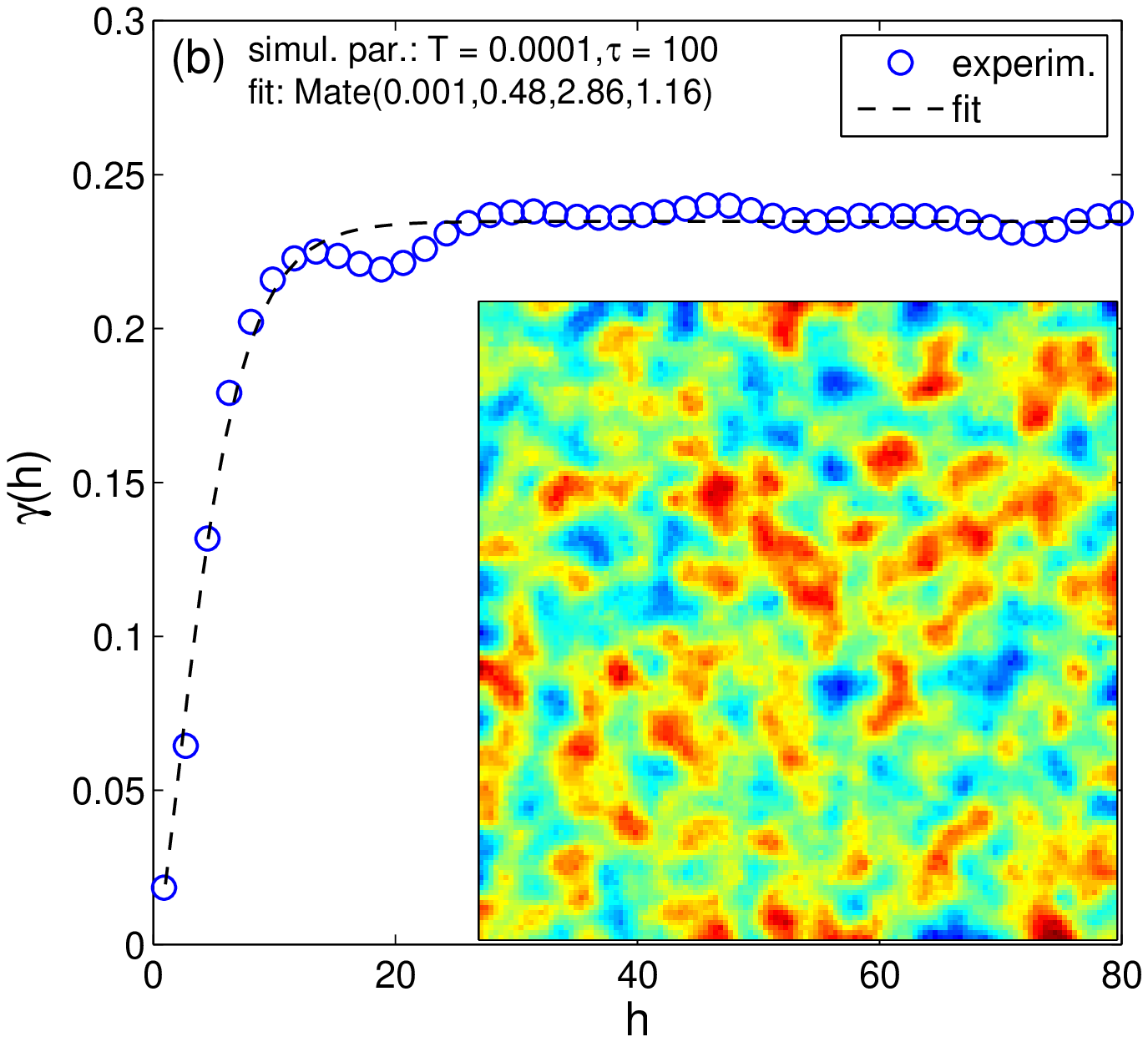}\label{fig:t00001_mcs100}}
\subfigure{\includegraphics[scale=0.54,clip]{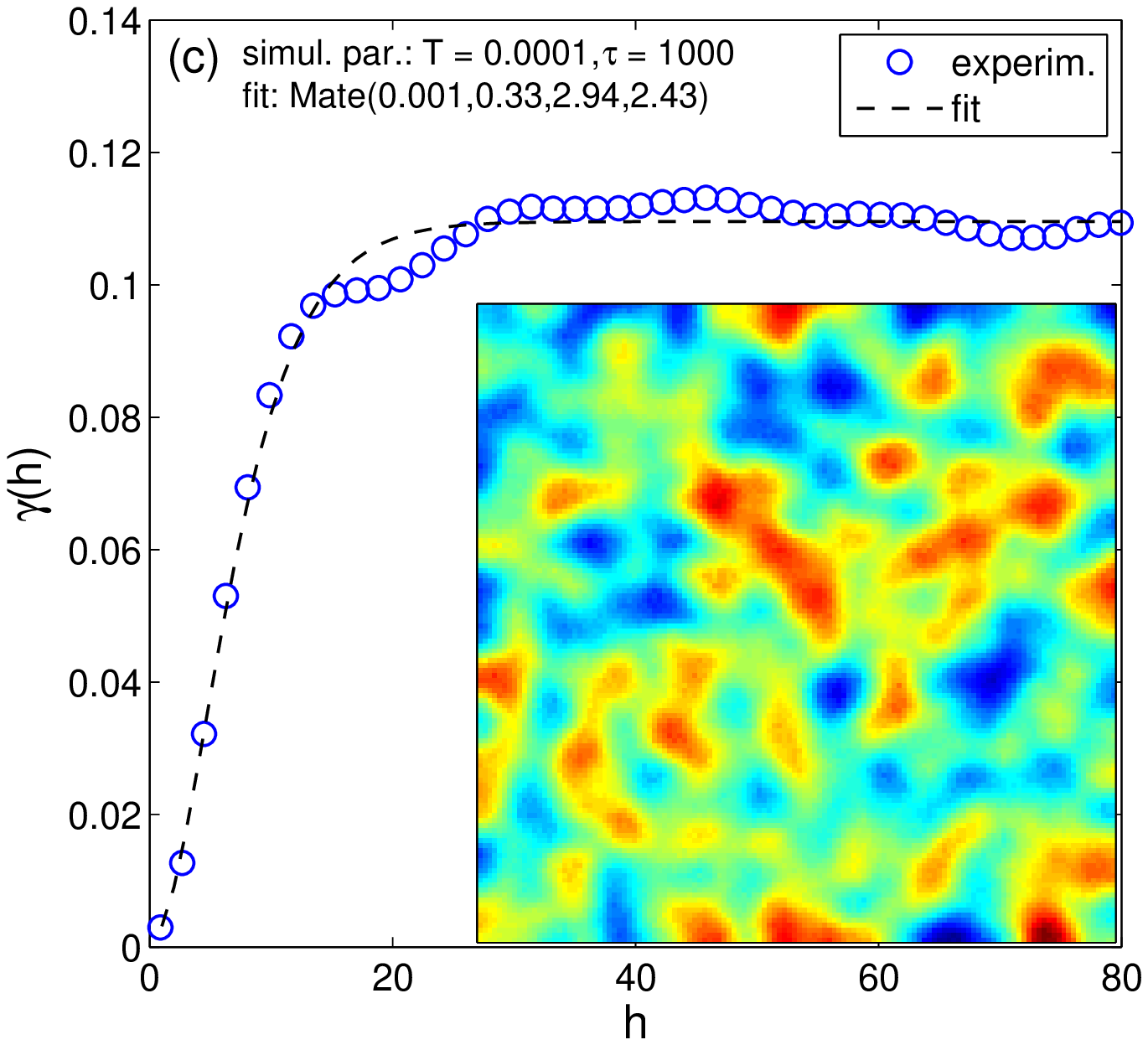}\label{fig:t00001_mcs1000}}
\subfigure{\includegraphics[scale=0.54,clip]{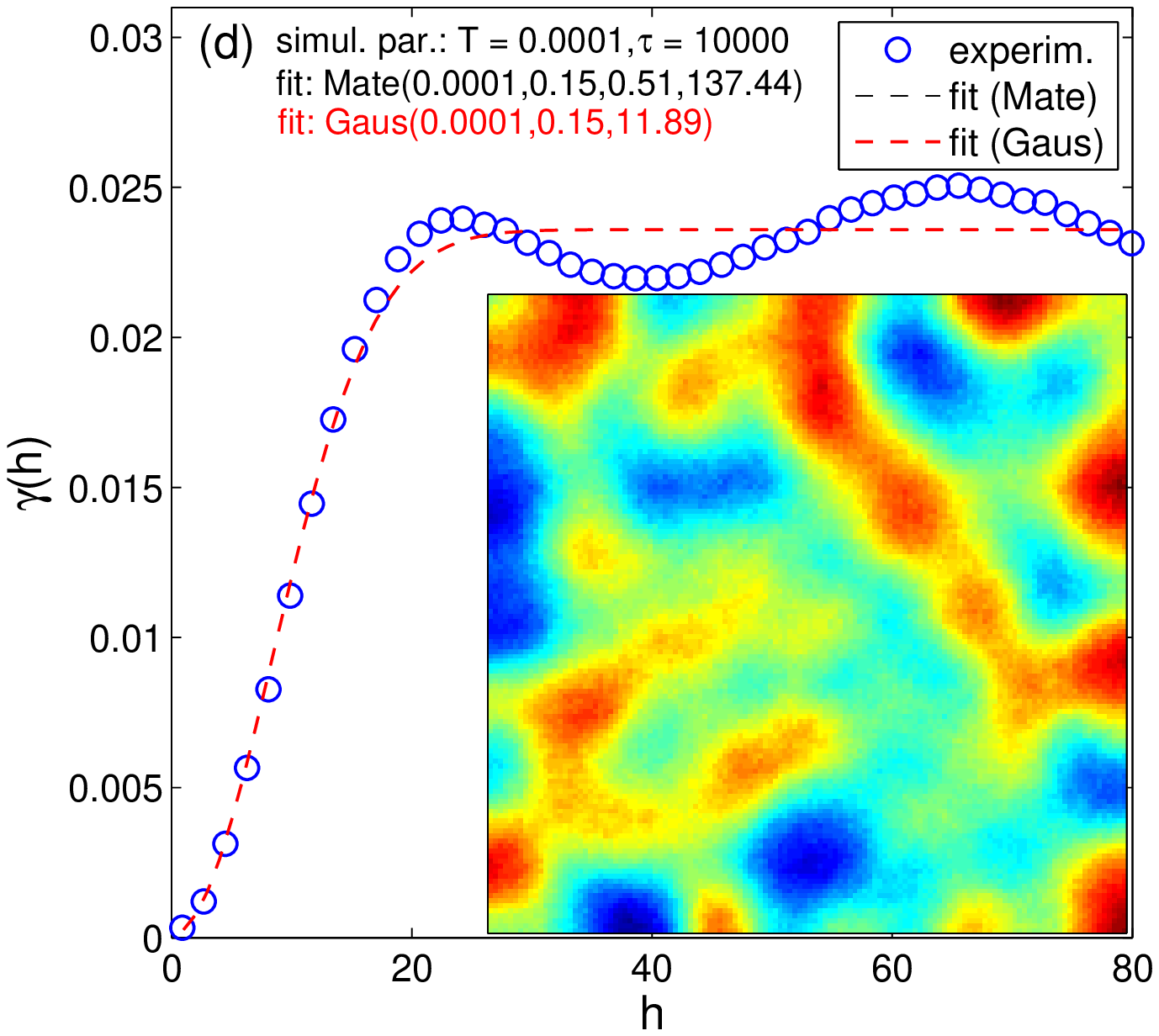}\label{fig:t00001_mcs10000}}
\caption{\label{fig:T00001}The same as in Fig.~\ref{fig:T001} for $T=10^{-4}$. Expo($\hat{\bm \vartheta}$) and Gaus($\hat{\bm \vartheta}$) denote respectively the exponential and Gaussian models for the inferred parameter values $\hat{\bm \vartheta}=(\hat{\sigma}_n,\hat{\sigma},\hat{\xi})$. Note that in Fig.~\ref{fig:t00001_mcs10} the fitted Mat\'{e}rn and exponential models coincide as do in Fig.~\ref{fig:t00001_mcs10000} the Mat\'{e}rn and the Gaussian models.}
\end{figure}

As the temperature is decreased the trend of building up correlations in the equilibration process is further enhanced. Since thermal fluctuations are gradually suppressed the realizations become smoother. This tendency is apparent by visual comparison of the snapshots obtained at $T=10^{-4}$, shown in Fig.~\ref{fig:T00001} with those generated in the same stages of the relaxation at $T=10^{-2}$, shown in Fig.~\ref{fig:T001}. The difference between the inferred values of the smoothness parameter for the respective cases becomes evident particularly at later stages ($\tau=10^3$ and $10^4$). As mentioned above, the exponential covariance model with quite rough (non-differentiable) realizations and the Gaussian model with very smooth (infinitely differentiable) realizations can be obtained from the Mat\'{e}rn model as special cases for $\nu=0.5$ and $\nu \rightarrow \infty$, respectively. To emphasize the flexibility of the correlations developed in the relaxation process we also fit the relatively rough data for $\tau=10$ (Fig.~\ref{fig:t00001_mcs10}) to the exponential and the very smooth data for $\tau=10^4$ (Fig.~\ref{fig:t00001_mcs10000}) to the Gaussian models. One can see an excellent collapse of both curves with the best fits to the Mat\'{e}rn model. In spite of the very small value of $\hat{\xi}$ inferred in the Mat\'{e}rn model for $\tau=10^4$ (see the explanation below), the characteristic length is expected to increase with decreasing temperature and increasing simulation time. This expectation can be justified by the fact that the ground state of the MPR model (the equilibrium state at $T=0$ corresponding to the lowest energy), is the ferromagnetic state with all the spins aligned in the same direction. Thus, for $T \rightarrow 0$ one can expect that the correlation (characteristic) length will span the entire lattice and eventually go to infinity for $L \rightarrow \infty$.

\begin{figure}[t!]
\begin{center}
\includegraphics[scale=0.54,clip]{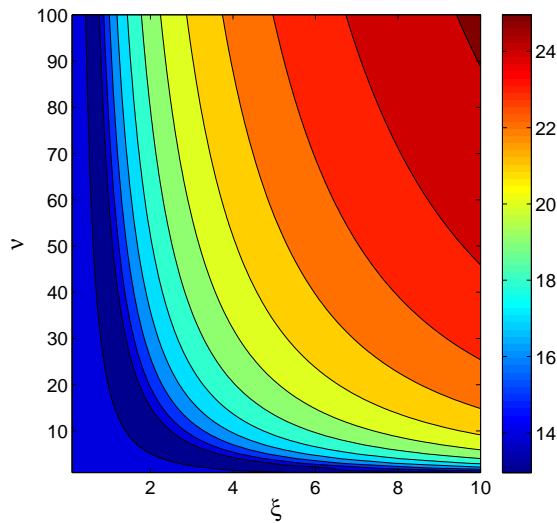}
\end{center}
\caption{\label{obj_fun}The objective function $O$ given by \eqref{wls} over the $\xi-\nu$ parameter space. $O$ represents the fit of the data obtained at $T=10^{-4}$ and $\tau=10000$ (see Fig.~\ref{fig:t00001_mcs10000}) to the Mat\'{e}rn model with $\hat{\sigma}_n=0.001$ and $\hat{\sigma}=0.15$.}
\end{figure}

Next, we comment on the  big discrepancy between the characteristic length parameters $\hat{\xi}$, estimated for the Mat\'{e}rn and Gaussian models, for $T=10^{-4}$ and $\tau=10^4$. In fact, in the Mat\'{e}rn model the parameter $\xi$ has been found to be highly negatively correlated with the smoothness parameter $\nu$~\cite{Minasny07,zuk09c,hris11}. This correlation is also apparent by looking at the objective function of the WLS fit, defined by Eq.~(\ref{wls}), which is shown in Fig.~\ref{obj_fun} in the $\xi-\nu$ parameter space, for the data presented in Fig.~\ref{fig:t00001_mcs10000}. It demonstrates that the samples with small $\hat{\xi}$ and large $\hat{\nu}$ values are quite ``close'' to the samples with large $\hat{\xi}$ and small $\hat{\nu}$ values. Put differently, a small difference between samples from the same population (particularly in non-ergodic samples) can lead to completely different parameter estimates. Thus, one should be careful when performing parameter inference from the empirical variogram. Recently, we have presented a model-independent definition of the correlation length borrowed from statistical field theory and proposed to use a so-called ergodicity index to compare coarse-grained measures corresponding to both trivial (standard) and non-trivial, e.g. Mat\'{e}rn, covariance models with different parameters~\cite{hris11}.

\section{Summary and outlook}
We have modified the standard planar rotator spin model from statistical physics to display a flexible type of short-range spatial correlations relevant in geostatistical modeling. In particular, the empirical study of spin configurations produced by Monte Carlo simulations in the nonequilibrium regime at various temperatures and stages shows that the smoothness and correlation range vary greatly versus the temperature and the simulation time. This behavior implies that the model has good potential for the simulation and prediction of spatial processes in geophysical and environmental applications. In particular, one can use the model to perform conditional simulations of gridded data, such as remote sensing images. Conditional simulations honor the sample values and reconstruct the variability of missing data, based on simulation parameters inferred from the available samples. Owing to the fact that the model does not show undesirable critical slowing down, the relaxation process is rather fast; furthermore, the short-range nature of the interactions between spin variables allows vectorization of the algorithm. Consequently, the proposed method is significantly more efficient than the conventional geostatistical approaches, and thus applicable to huge datasets, such as satellite and radar images. The implementation details are now under investigation.

\ack
This work was supported by the Scientific Grant Agency of Ministry of Education of Slovak Republic (Grant No. 1/0331/15). We also acknowledge support for a short visit by Prof. \v{Z}ukovi\v{c} at TUC from the Hellenic Ministry of Education - Department of Inter-University Relations, the State Scholarships Foundation of Greece and the Slovak Republic's Ministry of Education  through  the Bilateral Programme of  Educational Exchanges between Greece and Slovakia.

\end{document}